\documentclass[11pt]{article}
\usepackage{graphicx}
\usepackage{amssymb}
\usepackage{graphics}

\input{tcilatex}

\begin{document}

\title{A superfluid helium converter for accumulation and extraction of
ultracold neutrons}
\author{O. Zimmer$^{1,\ast }$, K. Baumann$^{1}$, M. Fertl$^{1}$, B. Franke$%
^{1}$, S. Mironov$^{1,2}$, C. Plonka$^{3}$, \and D. Rich$^{4,+}$, P.
Schmidt-Wellenburg$^{1,3}$, H.-F. Wirth$^{1}$, B. van den Brandt$%
^{5}\bigskip $ \\
$^{1}$Physik-Department E18, Technische Universit\"{a}t M\"{u}nchen, \\
D-85748 Garching, Germany\\
$^{2}$Laboratory of Nuclear Problems, JINR, Dubna, Moscow region 141980,
Russia\\
$^{3}$Institut Laue-Langevin, B.P. 156, 38042 Grenoble, France\\
$^{4}$Forschungsreaktor M\"{u}nchen FRM II, Lichtenbergstrasse 1, 85747
Garching, Germany\\
$^{5}$Paul Scherrer Institut, CH-5232 Villigen PSI, Switzerland}
\maketitle

\begin{abstract}
We report the first successful extraction of accumulated ultracold neutrons
(UCN) from a converter of superfluid helium, in which they were produced by
downscattering neutrons of a cold beam from the Munich research reactor.
Windowless UCN extraction is performed in vertical direction through a
mechanical cold valve. This prototype of a versatile UCN source is comprised
of a novel cryostat designed to keep the source portable and to allow for
rapid cooldown. We measured time constants for UCN storage and extraction
into a detector at room temperature, with the converter held at various
temperatures between $0.7$ and $1.3$ K. The UCN production rate inferred
from the count rate of extracted UCN is close to the theoretical
expectation.\bigskip


\bigskip

$^{\ast }$email: oliver.zimmer@ph.tum.de

$^{+}$new address: Oak Ridge National Laboratory, Tennessee, USA
\end{abstract}

\section{Introduction}

Ultracold neutrons (UCN) play an important role in fundamental
investigations in particle physics and cosmology. Searches for the neutron
electric dipole moment investigate CP-violation beyond the standard model of
particle physics \cite{Baker/2006,Pospelov/2005}. Accurate knowledge of the
neutron lifetime is required for understanding big bang nucleosynthesis \cite%
{Lopez/1999} as well as the semi-leptonic weak interaction within the first
quark family (see, e.g., contributions to a recent workshop in ref. \cite%
{Arif/2005}). Among other applications of UCN, the observation of quantum
states of the neutron in the gravitational field of the earth has attracted
recent interest \cite{Nesvizhevsky/2003}. The currently best source at the
Institut Laue-Langevin in Grenoble \cite{Steyerl/1986}\ provides UCN with
densities not exceeding a few $10$ per cm$^{3}$, which has motivated new
source projects in various places around the world \cite%
{Trinks/2000,Fomin/2000,Saunders/2004,Pokot/1995,Masuda/2002,Baker/2003,LANSCE/2004}%
.

As pointed out long ago, superfluid $^{4}$He can be used as a converter for
UCN production in a superthermal cooling process of cold neutrons \cite%
{Golub/1975}. As a result of the crossing dispersion relations of superfluid 
$^{4}$He and the free neutron, neutrons with wavelengths around $0.89$ nm,
i.e. $1.0$ meV kinetic energy, can be scattered down to the ultra-cold
energy range with emission of a single phonon. Multiphonon processes may
also contribute, depending on the neutron spectrum incident on the converter 
\cite{Korobkina/2002,Schott/2003}. Pure $^{4}$He has no neutron absorption
cross section, and at low temperature the density of excitations within the
helium is so small that upscattering of UCN back to higher energy becomes
unlikely. The UCN storage time $\tau $ may attain several $100$ s if the
converter vessel is made of a material with low UCN loss probability.
Ideally, it is surrounded with a magnetic trap as used in the neutron
lifetime experiment \cite{Huffman/2000}, where, for neutrons in the
low-field seeking spin state, $\tau $ may approach the neutron lifetime $%
\tau _{\mathrm{n}}=885.7\left( 8\right) $ s \cite{PDG/2006}. Past
experiments have already demonstrated UCN production rates in superfluid
helium close to the theoretical expectation \cite%
{Baker/2003,Huffman/2000,Ageron/1978}. It was concluded that, using an
intense cold neutron beam available at a high flux source one might realise
UCN densities up to several $10^{3}$ per cm$^{3}$. Vertical windowless
extraction of UCN from a superfluid helium bath and their subsequent
detection at room temperature was already demonstrated twice, first in an
early experiment performed at the ILL \cite{Ageron/1978}, and recently by a
Japanese group using a spallation neutron source \cite{Masuda/2002}.

The versatility of a superthermal helium UCN source would be strongly
improved if one could accumulate the UCN prior to their extraction, in order
to build up a high density. This was attempted $20$ years ago \cite%
{Kilvington/1987}, using a flap valve situated in the helium bath for
horizontal UCN extraction. However, the scheme failed to be efficient,
probably due to gaps and foils in the UCN transmission line, needed for
thermal protection of the helium converter. Although the measured rate of
upscattered neutrons was close to expectation, the rate of extracted UCN was
a factor of $50$ low. In turn, several groups decided to perform their
experiments within the superfluid helium without extracting them.

Here we report the first successful extraction of UCN from superfluid helium
after accumulation in the converter. Using a cold mechanical UCN valve
situated above the helium bath, no gaps or windows are required. The small
prototype involving a new type of cryostat enabled us to measure, with
negligible background, the UCN production rate and to study the
temperature-dependent storage properties of the converter. These first
experiments are very promising for versatile applications on a larger scale.

\section{Apparatus}

The central pieces of the apparatus are the UCN converter vessel with a cold
valve and connected tubing for UCN extraction (see fig. 1). The present
prototype has a rather small volume of about $2.4$ l. It is made from
electropolished stainless steel tubes (from the milk industry) with total
length $696$ mm and inner diameter $66$ mm and a neutron Fermi potential of $%
184\left( 4\right) $ neV. This defines, after subtraction of the Fermi
potential of the superfluid helium ($V_{\mathrm{F}}=18.5$ neV), the maximum
kinetic energy of storable neutrons. The incident cold neutron beam for UCN
production passes through two $0.125$ mm thick Ni foils ($V_{\mathrm{F}}=252$
neV) which close off the vessel on both sides. The valve for UCN extraction
is situated above the superfluid helium in a "T" section of the tube. It can
be manipulated from outside via a bellows-sealed capillary. With the valve
open, UCN may exit through a $7$ cm long vertical pipe with inner diameter $%
16$ mm. The subsequent extraction line consists of tapered transitions to
diameter $50$ mm, followed by a $90$ degree bend, then a horizontal $30$ cm
long straight guide, a conical section expanding to diameter $66$ mm, then
another $90$ degree bend, and finally a vertical $1$ m long straight section
down to a $^{3}$He gas UCN detector. All UCN guides are made from
electropolished stainless steel.%
\begin{figure}[h]\begin{center}
\includegraphics[
width=5.7938in]{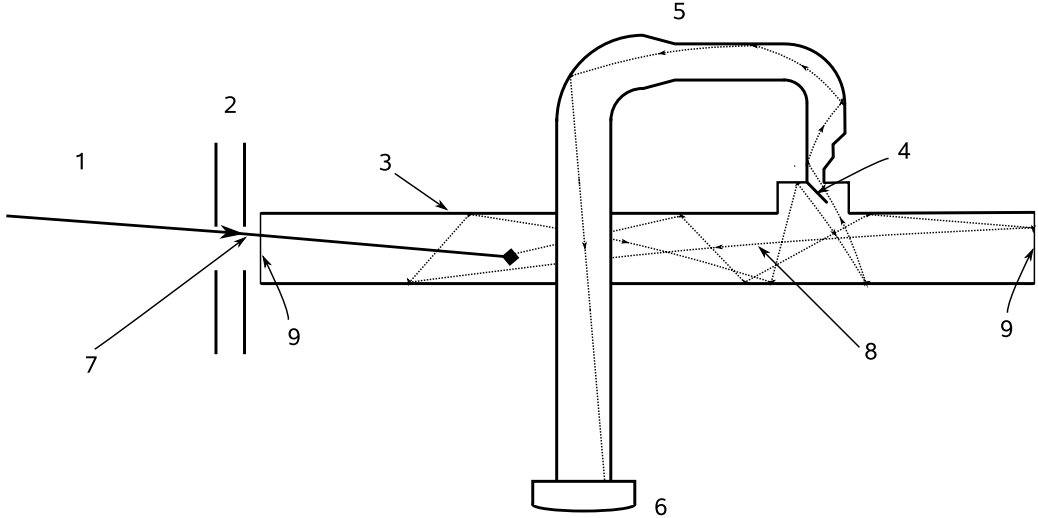}%
\caption{Schematics of the apparatus for UCN production in superfluid $^{4}$%
He. 1: cold neutron beam, 2: beam collimation, 3: stainless steel converter
vessel, 4: flap valve for UCN extraction, 5: UCN extraction guide, 6: UCN
detector, 7: cold neutron converted into a UCN, 8: trapped UCN, 9: Ni foils.}
\end{center}\end{figure}%

For filling and cooling the converter we developed a new cryostat. Primary
cooling power is provided by a commercial two-stage Gifford Mc-Mahon cold
head with a cooling power of $1.5$ W at $4.2$ K. It cools the thermal
radiation shields and liquefies helium from external gas supplies. Three
separate systems are thermally connected in cascade to the cold head: 1) a
continuous $^{4}$He evaporation stage to reach a temperature below the $%
\lambda $-transition to superfluidity, 2) a closed $^{3}$He system to reach $%
\lesssim 0.7$ K, and 3) a $^{4}$He filling line for the UCN converter. The
heat exchangers and condensers made from capillaries were already described
in ref.~\cite{Schmidt-Wellenburg/2006}. The $^{4}$He for the UCN converter
is supplied by a commercial gas cylinder, and is $99.999$ \% pure. To avoid
capture of UCN by residual $^{3}$He, the liquefied helium is passed through
a superleak held below the $\lambda $-transition temperature by the $^{4}$He
evaporation stage, which is supplied with liquefied helium through a needle
valve. The superleak consists of a stainless steel tube with inner diameter $%
7$ mm, filled with compressed Al$_{2}$O$_{3}$ powder with grain size $50$ nm
on a length of $15$ cm. The pure $^{4}$He then enters a heat exchanger
connected to the $^{3}$He evaporation stage. The interface is made from a
cylindrical copper disk with holes increasing the total surface to $200$ cm$%
^{2}$ on each side. Cooled close to the temperature of the liquid $^{3}$He,
the $^{4}$He flows to the converter vessel through a $16$ cm long tube with
inner diameter $1$ cm. Exploiting the high heat conductivity of the
superfluid, this results in a negligible temperature gradient. Using a roots
blower with $500$ m$^{3}$/h nominal pumping speed backed by a $40$ m$^{3}$/h
multiroots pump within the closed $^{3}$He cycle, we were able to cool the
filled converter down to $0.7$ K. The temperature was measured with a
calibrated cernox resistor attached to the converter volume. More details
about the cryostat will be published elsewhere.

\section{Experiments and results}

The apparatus was installed $1.7$ m behind the exit of the cold neutron
guide "NL1" at the Munich research reactor FRM II. The beam was collimated
from diameter $60$ mm down to diameter $33$ mm at the entrance to our
apparatus, thus defining a UCN production volume of $V_{\mathrm{p}}=595$ cm$%
^{3}$. The neutron particle flux density determined there by gold foil
activation was $1.5\times 10^{9}$ cm$^{-2}$s$^{-1}$ for a mean neutron
wavelength of $0.53$ nm.%
\begin{figure}[h]\begin{center}
\includegraphics[width=5.3299in]{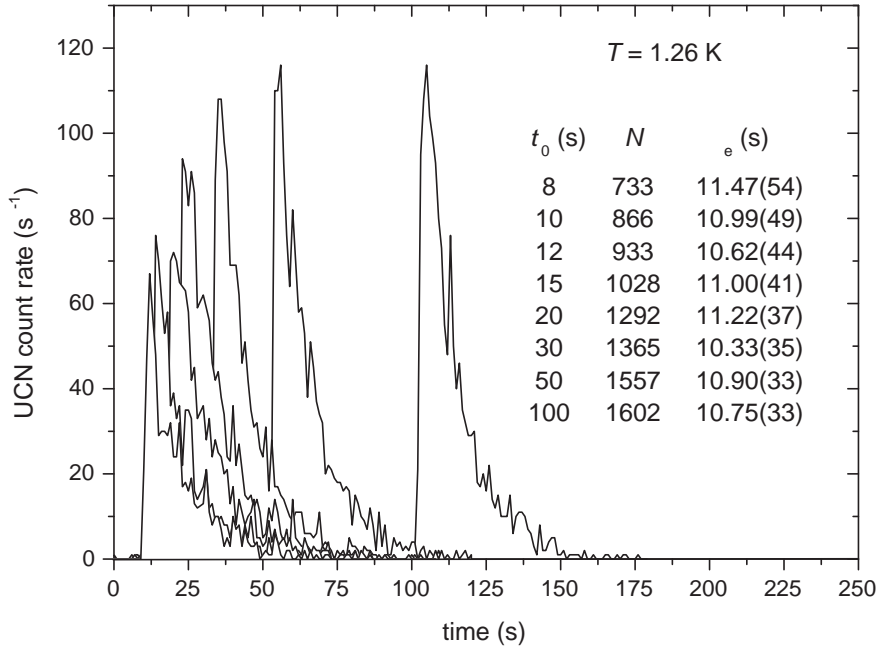}%
\caption{Time histograms of UCN count rates, measured in "buildup mode" at $%
1.26$ K for various UCN accumulation times $t_{0}$. Cold neutron irradiation
starts at $t=0$. $N$ denotes the integrated UCN counts for each of the
histograms and $\protect\tau _{\mathrm{e}}$ the corresponding emptying time
constant (see text).}
\end{center}\end{figure}%

In "buildup mode" measurements we first irradiated the converter with cold
neutrons with the UCN valve closed for an accumulation time $t_{0}$, after
which the beam was shut under simultaneous opening of the valve for UCN
detection. Figure 2 shows measured time histograms of UCN count rates at
temperature $T=1.26$ K. Their integrals, $N\left( t_{0}\right) $, follow
nicely a simple exponential saturation function $\propto 1-\exp \left(
-t_{0}/\tau \right) $. From the fit we obtained $\tau =\left( 13.28\pm
0.65\right) $ s, which is consistent with an earlier measurement at this
temperature, using a different method \cite{Golub/1983}. The time constant $%
\tau $ for UCN buildup is identical with the storage time constant at closed
UCN valve, which was also checked in a separate experiment. The rate $\tau
^{-1}$ has a $T$-independent but UCN energy-dependent contribution $\tau
_{0}^{-1}$ due to wall collisions, absorbing impurities, and UCN escape
through small holes in the vessel, and a $T$-dependent contribution $\tau _{%
\mathrm{up}}^{-1}$ due to UCN upscattering,%
\begin{equation}
\tau ^{-1}=\tau _{0}^{-1}\left( E\right) +\tau _{\mathrm{up}}^{-1}\left(
T\right) .
\end{equation}%
A second experimentally accessible quantity is the emptying time constant $%
\tau _{\mathrm{e}}$ deduced from the exponential decrease in each of the
histograms. It is related to $\tau $ and the time constant $\tau _{A}$ for
UCN passage through the extraction hole with area $A$ and the consecutive
guides by 
\begin{equation}
\tau _{\mathrm{e}}^{-1}\left( T,E\right) =\tau ^{-1}\left( T,E\right) +\tau
_{A}^{-1}\left( E\right) .
\end{equation}%
This follows from the proportionality of the detected rate to the UCN
density in the vessel, which decreases due to storage losses and due to UCN
extraction. At $T=1.26$ K, $\tau _{\mathrm{e}}$ was found to be independent
of $t_{0}$, with an average value $\tau _{\mathrm{e}}=\left( 10.82\pm
0.14\right) $ s. This is consistent with a single time constant $\tau $
being sufficient to describe UCN buildup in the vessel. These observations
tell us that, at this high temperature, $\tau ^{-1}$ is dominated by the UCN
energy-independent upscattering in the helium.%
\begin{figure}[h]\begin{center}
\includegraphics[
width=5.3299in]
{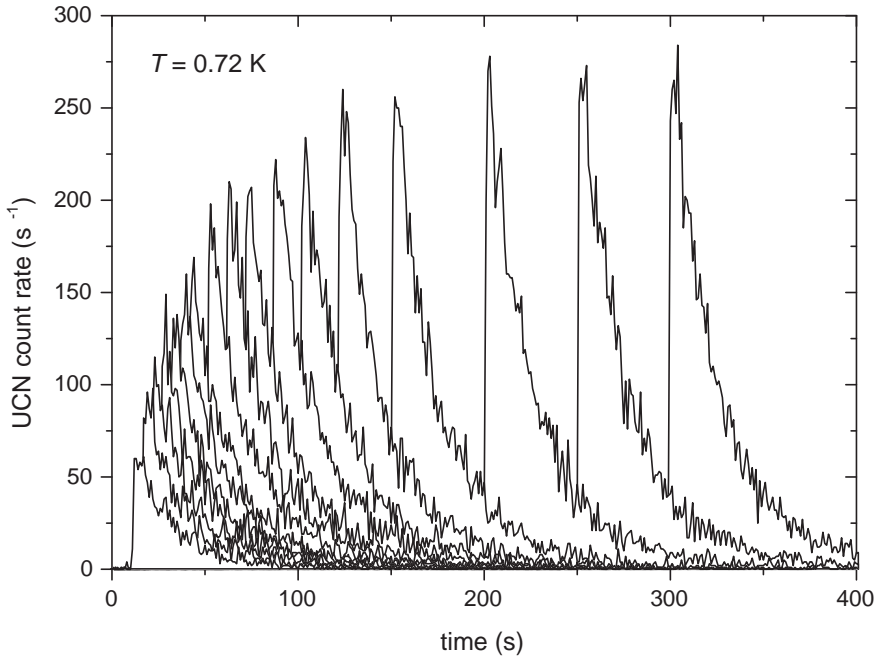}%
\caption{Time histograms of UCN count rates, measured in "buildup mode" at $%
0.72$ K for various accumulation times $t_{0}$. The corresponding emptying
time constants $\protect\tau _{\mathrm{e}}$ are shown in fig. 4. Note that
the UCN detector was always recording events starting from $t=0$. The few
counts for $0<t<7$ s are due to all measurements in this time interval,
which demonstrates the excellent background conditions.}
\end{center}\end{figure}%

Results of measurements at $0.72$ K show a much longer UCN buildup (see fig.
3 and compare with fig. 2). From a fit of the single-$\tau $ saturation
function to the integrals one obtains $\tau =\left( 55\pm 2\right) $ s.
Taking only the data for $t_{0}\geq 100$ s results in $\tau =\left( 69\pm
1\right) $ s. This demonstrates that a single exponential fit to the whole
data is no longer appropriate. An increase with accumulation time $t_{0}$ is
also observed for $\tau _{\mathrm{e}}$ (see fig. 4, where results for other
temperatures are also shown). The effects can be explained by UCN collisions
with the walls of the converter vessel: losses are more significant for
higher UCN energy (see, e.g., \cite{Golub/1991,Ignatovich/1990}), leading to
a reduction of the mean UCN velocity. This results in smaller losses due to
wall collisions and to slower extraction after longer accumulation.%
\begin{figure}[h]\begin{center}
\includegraphics[
width=5.3299in]
{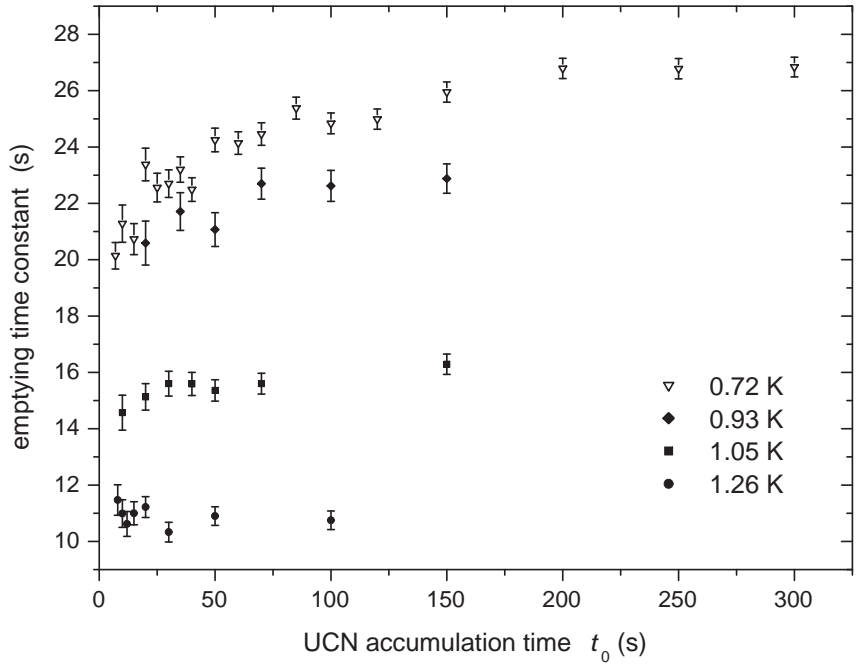}%
\caption{Emptying time constants $\protect\tau _{\mathrm{e}}$ as a function
of the UCN accumulation time $t_{0}$ in "storage mode" measurements for
various temperatures. Note that in the measurements at $0.93$ K the
converter vessel was only partly filled, leading to an increased $\protect%
\tau _{\mathrm{e}}$ and larger counting statistical uncertainties.}
\end{center}\end{figure}%

In "continuous mode" experiments we irradiated the converter with cold
neutrons while the UCN valve was open. At $1.26$ K, the UCN rate was $\dot{N}%
_{\mathrm{c}}=\left( 100\pm 5\right) $ s$^{-1}$. Defining $W=\varepsilon
\tau _{A}^{-1}/\tau _{\mathrm{e}}^{-1}=\left( 0.185\pm 0.041\right)
\varepsilon $ as the probability for a created UCN to become detected, with
the factor $\varepsilon $ accounting for imperfect detection efficiency and
losses in the UCN guide, we may determine the UCN production rate density%
\begin{equation}
P=\frac{\dot{N}_{\mathrm{c}}}{V_{\mathrm{p}}W}=\left( 0.91\pm 0.21\right)
/\varepsilon \text{ s}^{-1}\text{cm}^{-3}.  \label{P}
\end{equation}%
Due to the negligible count rate with closed UCN valve no background
correction of $\dot{N}_{\mathrm{c}}$ was needed.

The production rate density expected from the 1-phonon process in a helium
converter with Be wall coating ($V_{\mathrm{F}}=252$ neV) is $P_{\mathrm{I}%
}=\left( 4.55\pm 0.25\right) \times 10^{-9}\left. \text{d}\phi /\text{d}%
\lambda \right\vert _{\lambda ^{\ast }}$ s$^{-1}$cm$^{-3}$, with the
differential flux at $\lambda ^{\ast }=0.89$ nm given in cm$^{-2}$s$^{-1}$nm$%
^{-1}$ \cite{Baker/2003}. Using the results of prior time-of-flight (TOF)
measurements of the beam at the exit of NL1 \cite{Zeitelhack/2006} and
normalising the spectrum with our gold foil activation measurement of the
integral flux, we obtain $\left. \text{d}\phi /\text{d}\lambda \right\vert
_{\lambda ^{\ast }}=5.0\times 10^{8}$ cm$^{-2}$s$^{-1}$nm$^{-1}$. Including
a contribution of $30$ \% to UCN production due to multi-phonon processes in
addition to $P_{\mathrm{I}}$, as determined in ref.~\cite{Baker/2003} for a
cold beam with similar spectrum, and dividing by $1.68$ to account for the
reduction of UCN phase space due to the lower Fermi potential of stainless
steel, we might expect $P=1.76$ s$^{-1}$cm$^{-3}$. However, this value is
definitely an overestimation, as the preceeding analysis relies on earlier
TOF measurements, which were performed with a filling of the reactor's cold
source with $10.6$ liters liquid deuterium. In the present measurements, it
was operated with less than $9$ liters which leads to an intensity reduction
notably of the long-wavelength part of the spectrum (this effect is
described in ref.~\cite{Zeitelhack/2006}). In addition, divergence losses of 
$0.89$ nm neutrons in the beam collimation are more severe than for shorter
wavelengths. Both effects are not fully taken into account by normalisation
with our gold foil measurement. They are difficult to quantify without a
dedicated TOF measurement planned for future experiments. With this
uncertainty, our result for $P$ is indeed close to the expected value.

\section{Conclusion}

In summary, we have for the first time successfully extracted ultracold
neutrons accumulated in a converter of superfluid helium. Our setup provides
excellent background conditions which has enabled us to perform detailed
investigations of storage and emptying time constants. These studies will be
extended to lower temperatures and other storage materials in forthcoming
experiments. With an upgraded apparatus versatile applications for
experiments with UCN in vacuum are within reach. A specific application is
the measurement of the neutron lifetime using a magnetic trap, for which one
may also employ a different method of UCN extraction from the helium \cite%
{Zimmer/2005}.

\textbf{Acknowledgement} Thanks to Christian Hesse for his help with the
installation of equipment after midnight. This work is funded by BMBF under
contract number MT250.


\begin{thebibliography}{99}
\bibitem{Baker/2006} C.A. Baker et al., Phys. Rev. Lett \textbf{97} (2006)
131801.

\bibitem{Pospelov/2005} M. Pospelov, A. Ritz, Annals Phys. \textbf{318}
(2005) 119.

\bibitem{Lopez/1999} R.E. Lopez, M.S. Turner, Phys. Rev. D \textbf{59}
(1999) 103502.

\bibitem{Arif/2005} M. Arif et al., eds., Precision Measurements with Slow
Neutrons, J. Res. Nat. Stand. Technol. \textbf{110} (2005) 137.

\bibitem{Nesvizhevsky/2003} V.V. Nesvizhevsky, H.G. B\"{o}rner, A.K.
Petukhov, H. Abele et al., Nature \textbf{415} (2002) 297.

\bibitem{Steyerl/1986} A. Steyerl, H. Nagel, F.-X. Schreiber et al., Phys.
Lett. A \textbf{116} (1986) 347.

\bibitem{Trinks/2000} U. Trinks, F.J. Hartmann, S. Paul, W. Schott, Nucl.
Instr. Meth. A \textbf{440} (2000) 666.

\bibitem{Fomin/2000} A. Fomin et al., PSI Report TM-00-14-01 (2000), see
also http://ucn.web.psi.ch/.

\bibitem{Saunders/2004} A. Saunders, J.M. Anaya, T.J. Bowles et al., Phys.
Lett B \textbf{593} (2004) 55.

\bibitem{Pokot/1995} Y.N. Pokotilovski, Nucl. Instr. Meth A \textbf{356}
(1995) 412.

\bibitem{Masuda/2002} Y. Masuda, T. Kitagaki, K. Hatanaka et al., Phys. Rev.
Lett. \textbf{89} (2002) 284801-1.

\bibitem{Baker/2003} C.A. Baker, S.N. Balashov, J. Butterworth, P.
Geltenbort et al., Phys. Lett. A \textbf{308} (2003) 67.

\bibitem{LANSCE/2004} The LANSCE neutron edm experiment,
http://p25ext.lanl.gov/edm/edm.html.

\bibitem{Golub/1975} R. Golub and J.M. Pendlebury, Phys. Lett. \textbf{53A}
(1975) 133.

\bibitem{Korobkina/2002} E. Korobkina, R. Golub, B.W. Wehring, A.R. Young,
Phys. Lett. A \textbf{301} (2003) 462.

\bibitem{Schott/2003} W. Schott, J.M Pendlebury, I. Altarev, S. Gr\"{o}ger
et al., Eur. Phys. J. A \textbf{16} (2003) 599.

\bibitem{Huffman/2000} P.R. Huffman, C.R. Brome, J.S. Butterworth, K.J.
Coakley et al., Nature \textbf{403} (2000) 62.

\bibitem{PDG/2006} W.-M. Yao et al. (Particle data group), J Phys. G: Nucl.
Part. Phys. \textbf{33} (2006) 1.

\bibitem{Ageron/1978} P. Ageron, W. Mampe, R. Golub, J.M. Pendlebury, Phys.
Lett. \textbf{66A} (1978) 469.

\bibitem{Kilvington/1987} A.I. Kilvington et al., Phys. Lett. A \textbf{125}
(1987) 416.

\bibitem{Schmidt-Wellenburg/2006} P. Schmidt-Wellenburg and O. Zimmer,
Cryogenics \textbf{46} (2006) 799.

\bibitem{Golub/1983} R. Golub, C. Jewell, P. Ageron, W. Mampe et al., Z.
Phys. B \textbf{51} (1983) 187.

\bibitem{Zeitelhack/2006} K. Zeitelhack, C. Schanzer, A. Kastenm\"{u}ller et
al., Nucl. Instr. Meth. A \textbf{560} (2006) 444.

\bibitem{Golub/1991} R. Golub, D. Richardson, S.K. Lamoreaux, \textit{%
Ultra-Cold Neutrons}, IOP Publishing Ltd, 1991.

\bibitem{Ignatovich/1990} V.K. Ignatovich, \textit{The Physics of Ultracold
Neutrons}, Clarendon Press, Oxford 1990.

\bibitem{Zimmer/2005} O. Zimmer, Nucl. Instr. Meth. A \textbf{554} (2005)
363.
\end{thebibliography}
\end{document}